\documentclass[12pt,titlepage]{article}
\input amssym.def
\input amssym
\author{{\bf S.Leki\'c${}^1$, S.Galami\'c${}^2$ and Z.Rajili\'c${}^2$} \\
1)\,"Kosmos", Cetinjska 1, 78000 Banja Luka,\\ Republic of Srpska,
Bosnia and Herzegovina\\
2)\,Physics Department, Faculty of Science,\\ M.Stojanovi\'ca 2,
78000 Banja Luka,\\ Republic of Srpska, Bosnia and Herzegovina }
\title{Optical Fiber Communications: Group of the Nonlinear Transformations}

\begin{document}
\maketitle

\begin{abstract}
 A new method for finding solutions of the nonlinear Shr\"{o}dinger
equation is proposed. Comutative multiplicative group of the nonlinear
transformations, which operate on stationary localized solutions, enables
a consideration of fractal subspaces in the solution space,
stability and deterministic chaos. An increase of the
transmission rate at the optical fiber communications can be based on new
forms of localized stationary solutions, without significant change of
input power. The estimated transmission rate is $50\, Gbit/s$, for certain 
available soliton transmission systems.
\end{abstract}
\indent
$$ $$
$$ $$
$$ $$
\hspace{0.5cm} The propagation of pulsed light in an optical fiber can be described by the
nonlinear Schr\"{o}dinger equation,
$$i\frac{\partial q(\xi,\tau)}{\partial \xi}+\frac{1}{2}\frac{{\partial}^2
q(\xi,\tau)}{\partial \tau ^2}+{\mid q(\xi,\tau) \mid}^2q(\xi,\tau)=0,
\eqno (1)$$
where $q(\xi,\tau)$ is a complex envelope function of the effective
electric field amplitude and
$$\xi \propto x,\qquad \tau \propto (t - x\frac{\partial k}{\partial
\omega}). \eqno (2)$$
The higher order dispersion and the effect of  fiber loss are neglected
here \cite{[1]}. We take
$$q=q_0e^{i\frac{q_0^2}{2}\xi}y(\tau), \eqno (3)$$
where $y(\tau)$ is a real function, and get
$$y - \frac{1}{{q_0}^2}\frac{d^2y}{d\tau^2} - 2y^3 = 0. \eqno (4)$$
The solution of this equation \cite{[2]}
$$y_0(\tau)=\frac{1}{\cosh q_0\tau} \eqno (5)$$
describes the optical soliton. Its unchangeable shape is a property that
makes it attractive for applaying to ultra high speed optical
communications \cite{[3],[4]}. The equation (1) is completely integrable one.
The inverse scattering transformation method \cite{[5]} yields the general
solutions of such nonlinear partial differential equations. Our aim is to
propose here an alternative approach to the nonlinear Schr\"{o}dinger
equation and discus applicability of the obtained results to optical fiber
communications.\\

The equation (4) describes a stationary pulse in optical fiber. We take
a localized solution $y(\tau)$ of this equation and define the nonlinear
operator $H_{c_1}$:
$$H_{c_1}y = \sum_{j=1}^{\infty}c_j y^j , \eqno (6)$$
where $c_j$ are real coefficients. Does $H_{c_1}y$ satisfy the equation (4)?
The case $y=y_0$ is considered yet and the answer is positive \cite{[6]}. Putting
$H_{c_1}y$ into the equation (4), we find that $H_{c_1}y$ is actualy a
solution of this equation if
$$c_{2j}=0, \eqno (7)$$
while $c_{2j+1}$ satisfy the recursive relation
$$c_{2j+1}=\frac{1}{2j(j+1)} \{ j(2j-1)c_{2j-1} -
\sum_{n=2}^{2j}c_{2j+1-n}\sum_{k=1}^{n-1}c_{n-k}c_k \}, \eqno (8)$$
where $c_1$ is an arbitrary coefficient. Using the relations (6)-(8),
with $c_1=1$, we get
$$H_1 y=y. \eqno (9)$$
In the following text, $H_{c_1}$ will mean both the series (6) and the
recursion (8) with (7). For a localized $y(\tau)$ and a finite $c_1$,
convergence of the series (6) can be numerically tested. Our calculations
yield that $H_{c_1}y$ is localized too. Therefore, using different values
of $c_1$, we are able to get uncountable many new localized solutions of
the equation (4) from only one known localized solution (fig. 1). In the
following text "the solution" will mean "the localized solution of the
equation (4)". The solution value preciseness will be limited only by the
number of calculated coeficients. The solution in form different from (6) does not exist. Each
solution pair $z(\tau)$ and $y(\tau)$ must be in a relation $z=H_{c_1}y$,
with specific value of $c_1$:
$$c_1=\lim_{\tau \to \pm \infty} \frac{z(\tau)}{y(\tau)} .
\eqno (10)$$
Starting with a solution $y(\tau)$ we can construct the complete
solution space.
There is an analogy to the superposition principle from linear theory.
According to the relation (10), a solution is determined by its
asymptotics.\\

The nonlinear Schr\"{o}dinger equation has infinite number of symmetries
corresponding to the conserved quantities: total energy, momentum,
Hamiltonian, ... \cite{[1]}. We find that there are actually uncountable
conserved quantities. Let us consider the total energy only (for $H_{c_1}y$):
$$q_0^2\int_{-\infty}^{\infty} (c_1y+c_3y^3+c_5y^5+ ...)d\tau. \eqno (11)$$
We can choose uncountable different values of $c_1$ and use the relations
(7) and (8).\\

The relations (6)-(8) yield
$$H_{a_1}H_{b_1} = H_{a_1b_1} . \eqno (12)$$
Hence
$$\{ H_{c_1}; c_1 \not=0 \} \eqno (13)$$
is the comutative multiplicative group of the nonlinear transformations
(GNT). Group properties of the GNT originate from group properties of real
numbers $c_1 \not=0$. For example,
$$H_{c_1}H_{1/c_1}=H_1.\eqno (14) $$
For definite coefficient $c_1$ and solution $y(\tau)$, we can construct a
fractal subspace of the solution space. The fractal subspace covers
solutions of form 
$$H_{c_1}H_{c_1}...H_{c_1}y. \eqno (15)$$
In the phase plain, a fractal subspace is represented by a geometrical
fractal (fig. 2).

For optical fiber communications it is important question whether small
disturbations will destroy the information carrying pulses. Solution
parameters, amplitude (pulse width) and velocity (frequency), are affected
by various perturbations: outside produced noise, incoherence of the light
source, fiber inhomogenities, absorption, amplifier noise, soliton
interactions... It is the experimental fact that optical solitons
(equation (5)) are unlikely to be destroyed by perturbations - they are
very robust. We expect that at least the part of new solutions we have
expresed here are actually stable.
We are going to consider this problem theoretically, although it will be
open until an experimental verification.
The GNT method enables the following
statement: the stability of a solution $y(\tau)$ is equivalent to the
relation
$$\lim_{\epsilon \to 0} H_{1+\epsilon}y = y . \eqno (16)$$
The relations (6)-(8) and (16) yield
$$\mid y(\tau) \mid \leq 1. \eqno (17)$$
A localized solution of the equation (4) is stable one if and only if the
relation (17) holds (fig. 1a,b). As well as for the KdV soliton \cite{[7]}, the
classical argument about the counterbalance between nonlinearity and
dispersion is not sufficient to explain the stability.
Consideration of the Lyapunov exponent,
$$\lambda (c_1)=\lim_{j \rightarrow \infty} \frac{1}{j}ln \mid
\frac{dc_{2j+1}}{dc_1} \mid , \eqno (18) $$
yields that deterministic chaos will appear at close packing of solitons,
when $c_1$ is large enough (fig. 3a). The deterministic chaos we can
expect for $c_1 > 2.4$. Near $c_1=1$, stability is exceptional (fig. 3b).\\

New forms of localized stationary solutions of the nonlinear Schr\"{o}dinger 
equation enable an increase of the transmission rate at the optical fiber
communications, without significant change of input power. 
An information may be contained in the special form of 
soliton (fig. 1a,b). The known optical soliton, described by (5), is one of
many possible stationary pulses. Let us consider an available soliton 
transmission system. If the fiber core cross sectional area is $S=60 \mu m^2$,
the carrier wavelength is $\lambda =1.55 \mu m$, the soliton pulse (equation
(5)) width is $\tau_s=25ps$, the peak power is $P_m=2.1 mW$, and the
separation
between two adjacent solitons is $3\tau_s$, then the transmission rate is 
$10 Gbit/s$[8]. In the same transmission system, using stable pulses of form
$H_{c_1}y$ with different $c_1$ (fig. 1a), the transmission rate will
be greater.
It becomes equal to $50 Gbit/s$, at 40 photons resolution of energies.
The new forms of stable solutions (fig. 1b) make possible increase of
the transmission rate in the same system.\\

In conclusion, we have proposed the GNT method for solving of the
nonlinear Schr\"{o}dinger equation. New forms of the stationary
localized solutions, usable for an improvement of the optical fiber
communications, are obtained.\\

The authors would like to thank H.J.S.Dorren for useful discussion of the
GNT method. This work was supported by the Soros Fund Open Society
(Bosnia and Herzegovina) and the World University Service (Austria).

\newpage
\begin{figure}
{(a) Stable solutions with $c_1=0.2,\, 0.4,\, 0.6,\,
0.8,\, 1.0$ }  \\
{(b)Stable solutions with $c_1=1.4,\, 2.0,\, 2.39$.}  \\
\caption{(c)Unstable solutions with $c_1=2.41,\, 2.42$. }
\end{figure}

\begin{figure}
{2a}   \\
{2b}  \\
\caption{Phase diagrams of the solutions $(H_{c_1})^ny_0$. (a)$c_1=0.6,\, n=1$ to 8,
(b)$c_1=0.95,\, n=1$ to 5,  (c)$c_1=1.2,\, n=1$ to 5.}
\end{figure}

\begin{figure}
{(a)Lyapunov exponent}   \\
\caption{(b)Exceptional stability near $c_1=1$.}
\end{figure}

\end{document}